\documentclass[showpacs, pra, aps, superscriptaddress,twocolumn,10pt]{revtex4}
\usepackage{mathrsfs}
\usepackage{array}
\usepackage{amsmath}
\usepackage{graphicx}
\usepackage{amstext}
\usepackage{bm}
\usepackage{amsfonts}

\begin{document}
\title{Entangling two distant nanocavities via a waveguide}
\author{Hua-Tang Tan}
\affiliation{Department of Physics and Center for Quantum
information Science, National Cheng Kung University, Tainan 70101,
Taiwan} \affiliation{Department of Physics,
Huazhong Normal University, Wuhan 430079, China}
\author{Wei-Min Zhang}
\email{wzhang@mail.ncku.edu.tw} \affiliation{Department of Physics
and Center for Quantum information Science, National Cheng Kung
University, Tainan 70101, Taiwan}
\author{Gao-xiang Li}
\affiliation{Department of Physics, Huazhong Normal
University, Wuhan 430079, China}

\begin{abstract}
In this paper, we investigate the generation of continuous variable
entanglement between two spatially-separate nanocavities mediated by
a coupled resonator optical waveguide in photonic crystals. By
solving the exact dynamics of the cavity system coupled to the
waveguide, the entanglement and purity of the two-mode cavity state
are discussed in detail for the initially separated squeezing
inputs. It is found that the stable and pure entangled state of the
two distant nanocavities can be achieved with the requirement of
only a weak cavity-waveguide coupling when the cavities are resonant
with the band center of the waveguide. The strong couplings between
the cavities and the waveguide lead to the entanglement sudden death
and sudden birth. When the frequencies of the cavities lie outside
the band of the waveguide, the waveguide-induced cross frequency
shift between the cavities can optimize the achievable entanglement.
It is also shown that the entanglement can be easily manipulated
through the changes of the cavity frequencies within the waveguide
band.
\end{abstract}

\maketitle
\section{Introduction}
Entanglement of electromagnetic field with continuous variables
\cite{Braunstein1}, which embodies quantum correlations in amplitude
and phase quadratures of the field, has been proven to be very
important for building up continuous variable quantum communication
network \cite{Braunstein3,jia,Che09214538,shen} and quantum
computation \cite{oneway, error, gate}. Generally, continuous
variable entanglement is generated via nondegenerate optical
parametric oscillation \cite{jia, shen, lett, guman, li} or beam
splitters incorporating with squeezed light beams \cite{Braunstein3,
ch}.

Based on recent achievements in the light manipulation with highly
controllable photonic crystals \cite{notomi1}, we investigate in
this paper the generation of continuous variable entanglement
between two spatially-separated nanocavities mediated by a
coupled-resonator optical waveguide (CROW) in photonic crystals. As
we know, the long-distance entanglement is an essential ingredient
for transmitting quantum information over long distances in quantum
communication networks \cite{duan, kimble}. Very recently, the
distant entanglement between two qubits mediated by a plasmic
waveguide \cite{tudela} or two vibrating trapped ions assisted by a
phononic reservoir \cite{wolf} has just been studied. Here we shall
focus on the distant continuous variable entanglement between two
nanocavity fields. The main merits of the present scheme are as
follows. The nanocavity could be a point defect created in photonic
crystals and its frequency can be simply controlled just by changing
the size or the shape of the defect \cite{noda}. While the
waveguide, as the mediator of the two cavities as well as the output
channel, can be considered as a set of linearly coupled defects in
photonic crystals in which light propagates due to the coupling
between the adjacent defects \cite{yariv, notomi2}. The transmission
properties of the CROW can also be easily manipulated by changing
the modes of the resonators and the coupling configuration
\cite{baba}. Furthermore, the coupling of the cavity to the
waveguide is also controllable through the change of the distance
between the corresponding defects \cite{fan}. Besides the above
flexible controllability, the possibility of miniaturizing the
entanglement setup with solid-state photonic structures is highly
desirable for scalable and on-chip photonic quantum information
processing \cite{brien, linda}. Obviously, the bulk setups for
producing continuous variable entangled light in
Refs.\cite{Braunstein3,jia,shen,oneway, error, gate, lett, guman,
li, ch} are not suitable for integration. In addition, compared to
qubit or phononic entanglement proposed in Refs.\cite{tudela, wolf},
optical entanglement with continuous variables can be easily
detected and manipulated experimentally, which can make it very
promising for the implementation of continuous variable quantum
protocols \cite{Braunstein3, jia, Che09214538, shen, oneway, error,
gate}.

The effects of various environments on continuous variable
entanglement of optical fields or harmonic oscillators have been
extensively investigated, where one is mainly interested in the
non-Markovian decoherence dynamics of the entanglement
\cite{bm1,An,so2,so3,so4}. For the present system of two
spatially-separated nanocavities mediated by a waveguide, the
waveguide actually serves as a structured reservoir which is highly
controllable in experiments. We therefore are interested in the
entanglement generation and its manipulation through the controls of
the waveguide as well as the nanocavity properties. We shall study
the exact entanglement dynamics solved from the exact master
equation of the nanocavity system coupled to the waveguides that we
developed very recently \cite{xiong,wu,tan,lei}. The temporal
behavior of the entanglement and purity of the two-mode cavity state
is discussed in detail for initially separated squeezing inputs. We
find that when the cavity frequencies are resonant with the band
center of the waveguide, a stable and pure entangled state of the
two distant cavities can be generated with a requirement of only a
very weak coupling between the cavities and the waveguide. It shows
that the strong cavity-waveguide coupling will lead to the
entanglement sudden death and sudden birth \cite{yu}. When the
frequencies of the cavities locate outside the waveguide band, the
cross frequency shift of the nanocavities induced by the waveguide
can optimize the achievable entanglement. In addition, it is also
shown that the entanglement can be easily manipulated by changing
the cavity frequencies within the waveguide band.

The rest of the paper is organized as follows. In Sec.~II, we
formulate the entanglement dynamics of two spatially-separated
nanocavities coupled to a waveguide within the framework of the
exact master equation and the correlation matrix, the later
determines the continuous variable entanglement of the two
nanocavities through the measure of logarithmic negativity. In
Sec.~III, the properties of the entanglement and the corresponding purity are
investigated in detail. Finally, a conclusion is given in Sec. IV.

\section{Hamiltonian and entanglement measure}
\begin{figure}
\centerline{\scalebox{0.32}{\includegraphics{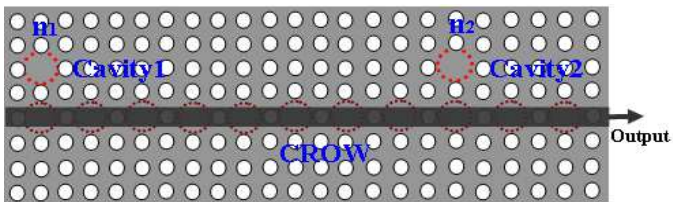}}} \caption{A
schematic plot of two spatially-separated nanocavities coupled to a
CROW structure at the sites $n_i~(i=1,2)$ in photonic crystals.}
\label{fig1}
\end{figure}

As schematically shown in Fig.~\ref{fig1}, we consider two
single-mode nanocavities coupled with a coupled resonator optical
waveguide (CROW) in photonic crystals at different sites. Each
cavity is formed by a point defect created in photonic crystals,
with the cavity frequency tunable by changing the geometrical
parameters of the defect. The waveguide in photonic crystals
consists of a series of coupled point defects in which light
propagates due to the coupling between the adjacent defects.
Experimentally, the large-scale CROW consisting of more than one
hundred coupled resonators has been successfully fabricated
\cite{noda}. To be specific, let the cavity 1 be coupled to the
waveguide at the site $n_1$ and the cavity 2 coupled to the
waveguide at the site $n_2$, as shown in Fig.~\ref{fig1}. By
treating the waveguide as a tight-binding model, the
Hamiltonian of the whole system is given by \cite{wu}
\begin{align}
H=& \sum_{i=1,2}\omega_i a_i^\dag a_i+\sum_{k}\omega_kb_k^\dag b_k
+\sum_{i,k}V_{ik}(a_ib_k^\dag+b_k a_i^\dag),\label{H0}
\end{align}
where
\begin{align}
\omega_k=\omega_0-2\xi_0\cos k,\  ~~
V_{ik}=\sqrt{\frac{2}{\pi}}\xi_i\sin (n_ik),\label{kk}
\end{align}
with $0\le k\le \pi$. The operators $a_i$ and $a_i^\dag$ are the
annihilation and creation operators of the cavity fields with
frequencies $\omega_i$. The annihilation and creation operators
$b_k$ and $b_k^\dag$ describe the Bloch modes $\omega_k$ of the
waveguide with $\omega_0$ being the identical frequency of each
resonator in the waveguide. The frequencies $\omega_i$ and
$\omega_0$ are tunable by adjusting the geometrical parameters of
the corresponding defects. The strength $\xi_0$
characterizes the photon hopping between two adjacent resonators in
the waveguide and is controllable by changing the corresponding distance between
the two defects. The controllable coupling strength $\xi_i$ is the
coupling of the $i$th cavity to the waveguide at the sites
$n_i$. We should point out that the frequencies of the two cavities and
the waveguide band considered in the above system should lie inside
the photonic band gap of the photonic crystals. Then the photon loss
into the photonic crystals is totally negligible \cite{Notomi}.

We will investigate the generation of the entanglement between the
two spatially-separated cavity fields through the controllable
waveguide, where the two nanocavities are initially prepared in a
separated two-mode Gaussian state. For a two-mode Gaussian state,
its quantum statistical property is fully determined by the
$4\times4$ correlation matrix $\bm \chi$ which is defined by
\begin{align}
\label{cm} \chi_{ij}=\frac{1}{2}\langle x_ix_j+x_jx_i\rangle,
\end{align}
where the vector $x=(X_1, Y_1, X_2, Y_2)$ and the quadrature
operators $X_j=(a_j+a_j^\dag)/\sqrt{2}$ and
$Y_j=-i(a_j-a_j^\dag)/\sqrt{2}$. With the correlation matrix $\bm
\chi$, the continuous variable entanglement between the two cavity
fields can be well quantified with the measure of logarithmic
negativity. By re-expressing the correlation matrix $\bm\chi$ in
terms of three $2\times2$ matrices $\bm\varrho_1$, $\bm\varrho_2$,
and $\bm\varrho_3$,
\begin{align}
\bm\chi=\begin{pmatrix}  \bm\varrho_1 &  \bm\varrho_3 \\
 \bm\varrho_3^T& \bm\varrho_2 \end{pmatrix}, \label{sigm}
\end{align}
the logarithmic negativity $E_N(\bm\chi)$ is defined as \cite{vidal}
\begin{align}
\label{ln} E_N(\bm\chi)=\textrm {max}[0, -\textrm{ln}(2\lambda)],
\end{align}
where
\begin{align}
\lambda=\frac{\sqrt{\Delta-\sqrt{\Delta^2-4~\textrm{Det}\bm
\chi}}}{\sqrt{2}},
\end{align}
and $\Delta=\textrm{Det} \bm\varrho_1+\textrm{Det}
\bm\varrho_2-2~\textrm{Det} \bm\varrho_3$. Thus, the entanglement
between the two cavity fields occurs for $E_N(\bm\chi)>0$, i.e., for
$\lambda>\frac{1}{2}$. From the definitions of Eqs.~(\ref{cm}) and
(\ref{sigm}), we see that the matrix elements of $\bm\chi$  are a
linear function of the second-order quantities
\begin{align}
n_{ij}(t)=\langle a^\dag_i(t) a_j(t)\rangle,  ~~ s_{ij}(t)=\langle
a_i(t) a_j(t)\rangle \label{soq0}
\end{align} plus its
hermitian conjugate. The entanglement dynamics between the two
cavity fields at any time is then completely determined by these
time-dependent second-order quantities.

On the other hand, the Hamiltonian of Eq.~(\ref{H0}) describes
effectively the two spatially-separated nanocavities coupled to a
common reservoir with a controllable spectral structure. In other
words, the waveguide plays a role of a structured reservoir as a
mediator between two nanocavities. We can use the exact master
equation we developed recently for nanocavities coupled to
waveguides in photonic crystals \cite{wu,tan,lei} to investigate the
exact entanglement dynamics of the two nanocavities coupled to the
waveguide in photonic crystals. By assuming that two nanocavities
are initially uncorrelated to the structured reservoir (i.e., the
waveguide) and the waveguide is initially in vacuum, the exact
master equation for the density operator $\rho(t)$ of the cavity
system can be obtained through the Feynman-Vernon influence
functional approach \cite{Fey} in the framework of coherent state
path-integral representation \cite{Zhang2}. The resulting exact
master equation is given by
\begin{align}
&\frac{d\rho(t)}{dt}= -i[H_{\textrm{eff}}(t), \rho(t)] \notag \\
&~~+\sum_{ij}\gamma_{ij}(t)[2a_j\rho(t) a_i^\dag -a_i^\dag a_j
\rho(t)-\rho(t) a_i^\dag a_j],\label{me}
\end{align}
where $H_{\textrm{eff}}(t)=\omega_{ij}(t)a_i^\dag a_j$ is the
effective Hamiltonian of the cavity fields with the time-dependent
renormalized frequencies $\omega_{ij}(t)$, which is resulted from the
back-reaction of the the waveguide to the cavity system. The
time-dependent coefficients $\bm\gamma(t)$ describe the dissipation
of the cavity system due to the coupling to the waveguide. The
coefficients $\omega_{ij}(t)$ and $\bm\gamma(t)$ are
non-perturbatively determined by
\begin{subequations}
\label{tdcoefs}
\begin{align}
\omega_{ij}(t)&=\frac{i}{2}[\dot{\bm\mu}(t)\bm\mu^{-1}(t)-h. c.]_{ij},\\
\gamma_{ij}(t)&=-\frac{1}{2}[\dot{\bm\mu}(t)\bm\mu^{-1}(t)+h. c.]_{ij},
\end{align}
\end{subequations}
where $\bm\mu(t)$ is the cavity photon propagating function which
obeys the integrodifferential equation of motion
\begin{align}
\frac{d}{dt}\bm\mu(t)=-i\bar{\bm\omega}\bm\mu(t)-\int_{t_0}^t  \bm
g(t-\tau)\bm\mu(\tau)d\tau,\label{ut}
\end{align}
with the initial condition $\bm\mu(t_0)=1$.
Here, the frequency matrix $\bar{\bm\omega}=\textrm{Diag}[\omega_1,
\omega_2]$ is a diagonal frequency matrix of the two cavities. The
integral kernel in Eq.(\ref{ut}) involves the time-correlation
function $\bm g(\tau-\tau')$, which non-perturbatively characterizes
the non-Markovian memory structure between the cavity system and the
waveguide.

\begin{figure}
\centerline{\scalebox{0.33}{\includegraphics{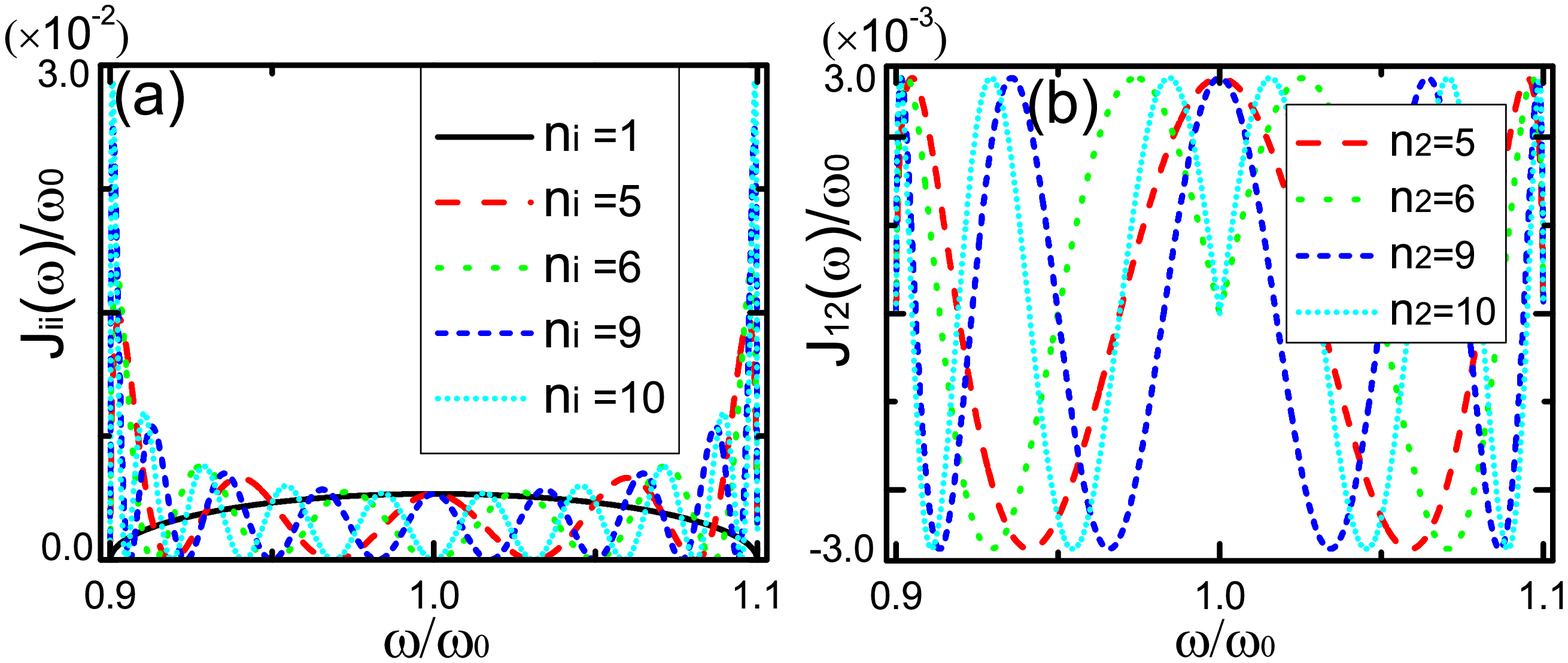}}} \caption{(a)
The spectral density $J_{ii}(\omega)$ for the cavities at different
locations $n_i$. (b) The spectral density $J_{12}(\omega)$ for the
different sites $n_2$ of the second cavity and the first cavity
location $n_1=1$.  Here the two cavity frequencies are taken to be
equal and also equal to the resonator frequency in the waveguides:
$\omega_1=\omega_2=\omega_0$. The cavity-waveguide coupling
$\xi_1=\xi_2=0.2\xi_0$, and the hopping rate in the waveguide
$\xi_0=0.05\omega_0$.} \label{fig2}
\end{figure}

By introducing the spectral densities of the waveguide:
$J_{ij}(\omega)=2\pi\sum_{k}V_{ik}V_{jk}\delta(\omega-\omega_k)$,
the time-correlation function can be expressed as
\begin{subequations}
\begin{align}
g_{ij}(\tau-\tau')&=\int \frac{d\omega}{2\pi}J_{ij}(\omega)e^{-i\omega(\tau-\tau')}.
\end{align}
\end{subequations}
Since the waveguide in photonic crystals has a narrow but continuous
band structure, the spectral densities becomes $J_{ij}(\omega)=2\pi
\varrho(\omega)V_i(\omega)V_j(\omega)$, where $\varrho(\omega)$ is
the density of states in the waveguide determined by Eq.~(\ref{kk}).
Explicitly, we have
 \begin{subequations}
\begin{align}
& \varrho(\omega)=\frac{1}{\sqrt{4\xi_0^2-(\omega-\omega_0)^2}},\\
V_i(\omega)=\sqrt{\frac{2}{\pi}}&
\xi_i\sin[n_i\arcsin(\frac{\sqrt{4\xi_0^2-(\omega-\omega_0)^2}}{2\xi_0})],
\end{align}
\end{subequations}
where $\omega_0-2\xi_0\leq\omega\leq\omega_0+2\xi_0$ is  the band of
the waveguide. In Fig.~\ref{fig2}, the spectral densities
$J_{11}(\omega)=J_{22}(\omega)$ and $J_{12}(\omega)=J_{21}(\omega)$
are plotted as a function of $\omega$ with different distances
between the two cavities, for the case of the equal cavity
frequencies $\omega_1=\omega_2=\omega_c$ and the equal
cavity-waveguide couplings $\xi_1=\xi_2=\xi_c$. As we will show
later in the next section, the entanglement characteristics depend
heavily on the spectral structures.
In fact, it is the back-action through the off-diagonal matrix
element $J_{12}(\omega)$ of the spectral density that induces an
effective coupling between the two separated nanocavities which
leads to the entanglement generation. The environment-assisted
continuous variable entanglement has been investigated in the
literature \cite{so2, s1, s2}. However, the entanglement in these
investigations is uncontrollable. Here, the waveguide (a structured
reservoir)-induced entanglement between two spatially-separated
nanocavities in photonic crystals is fully controllable and is
promising for quantum information processing in all-optical
circuits.

Now, the exact temporal behavior of the second-order quantities of
Eq.~(\ref{soq0}) can be completely determined by the exact master
equation. Explicitly, from the exact master equation (\ref{me}), it
is not too difficult to find that the second-order quantities obey
the following equations
\begin{subequations}
\begin{align}
\frac{d}{dt}\bm n(t)&=[\dot{\bm\mu}(t)\bm\mu^{-1}(t)]\bm n(t)+\bm
n(t)[\dot{\bm\mu}(t)\bm\mu^{-1}(t)]^\dag,\\
\frac{d}{dt}\bm s(t)&=[\dot{\bm\mu}(t)\bm\mu^{-1}(t)]\bm s(t)+\bm
s(t)[\dot{\bm\mu}(t)\bm\mu^{-1}(t)]^T.
\end{align}
\end{subequations}
The exact solutions of the above equations are found to be
\begin{align}
\bm n(t)=\bm \mu(t)\bm n(0)\bm\mu^\dag(t) ~,~~ \bm s(t)=\bm
\mu(t)\bm s(0)\bm\mu^T(t),\label{sec}
\end{align}
where $\bm n(0)$ and $\bm s(0)$ are the initial second-order
quantities. Thus, once the photon propagating function $\bm \mu(t)$
is solved from the integrodifferential equations of motion
(\ref{ut}), the above exact solution allows us investigate the
entanglement generation of the two separated nanocavities. Meantime,
the temporal evolution of the entanglement measure subjected to the
non-Markovian dissipation and fluctuation from the structured
reservoir can be fully taken into account.

\section{Results and discussion}
Now we are ready to investigate the entanglement generation between
the two cavity fields and its temporal evolution for a given
initially-separated state. We assume that the two cavity fields are
initially prepared in a single-mode squeezed state, respectively,
i.e.,
\begin{eqnarray}
|\psi_{a_i}(0)\rangle=\exp(-\frac{r_i}{2}a_i^2+\frac{r_i}{2}a_i^{\dag2})
|0_i\rangle,~~ i=1,2 \label{inista}
\end{eqnarray}
where the  squeezing parameter $r_i$ is controllable as an input.
The preparation of the initial squeezed states can be well
accomplished by injecting into the nanocavities squeezed radiation
fields produced via degenerate parametric oscillation \cite{b1}. For
the initial states of Eq.~(\ref{inista}), the initial average photon
numbers and two-photon correlations are given by
\begin{align}
 n_{ij}(0)=\sinh^2(r_i)\delta_{ij},~~
 s_{ij}(0)=\sinh(r_i)\cosh(r_i)\delta_{ij}.
\end{align}
With the help of the above initial conditions, we can analyze the
temporal behavior of the correlation matrix $\bm \chi$ in
Eq.~(\ref{cm}), and then the logarithmic negativity $E_N$ given by
Eq.~(\ref{ln}). In the following, we will discuss the entanglement
in three cases: (i) the cavity frequency $\omega_c$ is resonant with
the band center $\omega_0$ ($\omega_c=\omega_0$); (ii) the cavity
frequency $\omega_c$ stays the outside of the waveguide band
($|\omega_c-\omega_0|\ge2\xi_0$); and (iii) the cavity frequency
$\omega_c$ lies within the waveguide band
($|\omega_c-\omega_0|<2\xi_0$).

\subsection{$\omega_c=\omega_0$}
\begin{figure}
\centerline{\scalebox{0.3}{\includegraphics{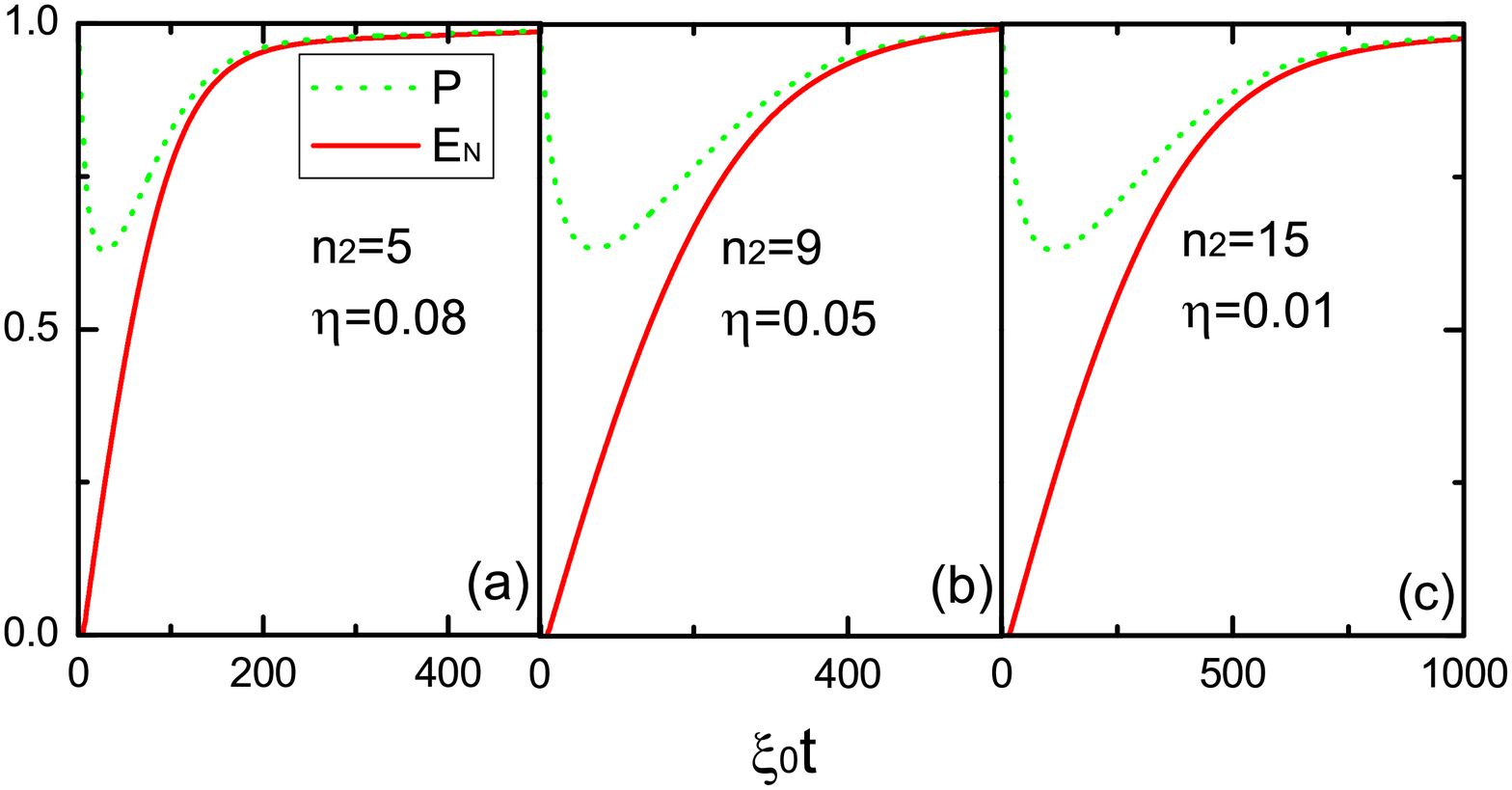}}} \caption{The
temporal behavior of the logarithmic negativity $E_N$  and the
entanglement state purity $P$ for different values of the coupling
$\eta$ and the site $n_2$ the second cavity located. The cavity
frequency $\omega_c=\omega_0$, the hopping rate
$\xi_0=0.05\omega_0$, the squeezing parameter $r=1.0$, and the site
of the first cavity $n_1=1$.} \label{fig3}
\end{figure}

At first, we consider the situation that the frequency $\omega_c$ of
the two cavities is resonant with the band center $\omega_0$ of the
waveguide. In this case, the steady-state average values in
Eq.~(\ref{sec}) in the weak cavity-waveguide coupling 
can be analytically obtained from
Eq.~(\ref{bm}) as
\begin{subequations}
\label{ssq}
\begin{align}
&\langle a_1^\dag a_1\rangle_s=\frac{\gamma_{22}\sinh^2r}{\gamma_{11}
+\gamma_{22}},~~~ \langle a_2^\dag a_2\rangle_s=\frac{\gamma_{11}
\sinh^2r}{\gamma_{11}+\gamma_{22}},\\
&\langle a_1^2\rangle_s=\frac{\gamma_{22}'\sinh2r}{2(\gamma_{11}
+\gamma_{22})},~~~ \langle a_2^2\rangle_s=\frac{\gamma_{11}'
\sinh2r}{2(\gamma_{11}+\gamma_{22})},\\
&\langle a_1 a_2\rangle_s=-\frac{\gamma_{12}'\sinh2r}{2(\gamma_{11}
+\gamma_{22})},~~\langle a_1^\dag a_2\rangle_s=-\frac{\gamma_{12}
\sinh^2r}{\gamma_{11}+\gamma_{22}}, \label{ssqc}
\end{align}
\end{subequations}
where the subscript "s" denotes the steady state, $\gamma_{ii'}'
=\gamma_{ii'}\textrm e^{-2i\omega_ct}$, and the initial squeezing
$r_1=r_2=r$ is also assumed. From the above results, the expression
of the stationary logarithmic negativity $E_N^s$ can be obtained and
the result is rather cumbersome so that we do not give it here
explicitly. Nevertheless, it can be seen from Eq.~(\ref{ssq}) that
the waveguide-induced collective effect between the two cavities
(characterized by the cross damping factor $\gamma_{12}$ in
Eq.~(\ref{ssqc})) is crucial for generating the entanglement. Since
the cross damping rate $\gamma_{12}=\sqrt{\gamma_{11}\gamma_{22}}$
with $\gamma_{ii}=(\frac{2\xi_i^2}{\xi_0})\sin^2(\frac{n_i\pi}{2})$
in the weak coupling region, the stationary entanglement cannot be
generated if any one (or both) of the sites $n_1$ and $n_2$ of the
two cavities is even. Only when $n_1$ and $n_2$ are both odd
numbers, can the cross damping rate not vanish. For the equal
couplings $\xi_1=\xi_2=\xi_c$, we have
$\gamma_{11}=\gamma_{22}=2\xi_c^2/\xi_0$. Then, the steady-state
logarithmic negativity $E_N^s$ is reduced to
\begin{align}
E_N^s=r.
\end{align}
This shows that the entanglement degree between the two
spatially-separated nanocavities can be controlled by the input
initial squeezing parameter in the resonant case. Furthermore, from
the definition of the purity $P=\textrm
{Tr}(\rho^2)=\frac{1}{4\sqrt{\textrm{Det}\bm \chi}}$, it is also not
difficult to find that the steady purity $P_s=1$ for the steady
state of Eq.~(\ref{ssq}). This can be clearly understood  by
performing an unbalanced beam-splitter transformation
$d_{1}=\sin\theta a_1+\cos\theta a_2$ and $d_{2}=\cos\theta
a_1-\sin\theta a_2$ on the Hamiltonian in Eq.~(\ref{H0}), where
$\sin\theta=V_{1k}/\sqrt{V_{1k}^2+V_{2k}^2}$. Then the whole
Hamiltonian of Eq.~(\ref{H0}) is reduced to $H=\sum_i\omega_c
d_i^\dag d_i+\sqrt{V_{1k}^2+V_{2k}^2}(d_1b_k^\dag+b_k d_1^\dag)$. It
shows that the waveguide only couples to the collective mode $d_1$,
while the other collective mode $d_2$ is totally decoupled from the
waveguide. Therefore, the collective mode $d_2$ of the two cavities
forms a decoherence-free subspace, which does not subject to any
dissipation due to the presence of the structured reservoir. This
results in the steady and pure entangled state between the two
distant nanocavities, as shown in Fig.~\ref{fig3}
\begin{figure}
\centerline{\scalebox{0.3}{\includegraphics{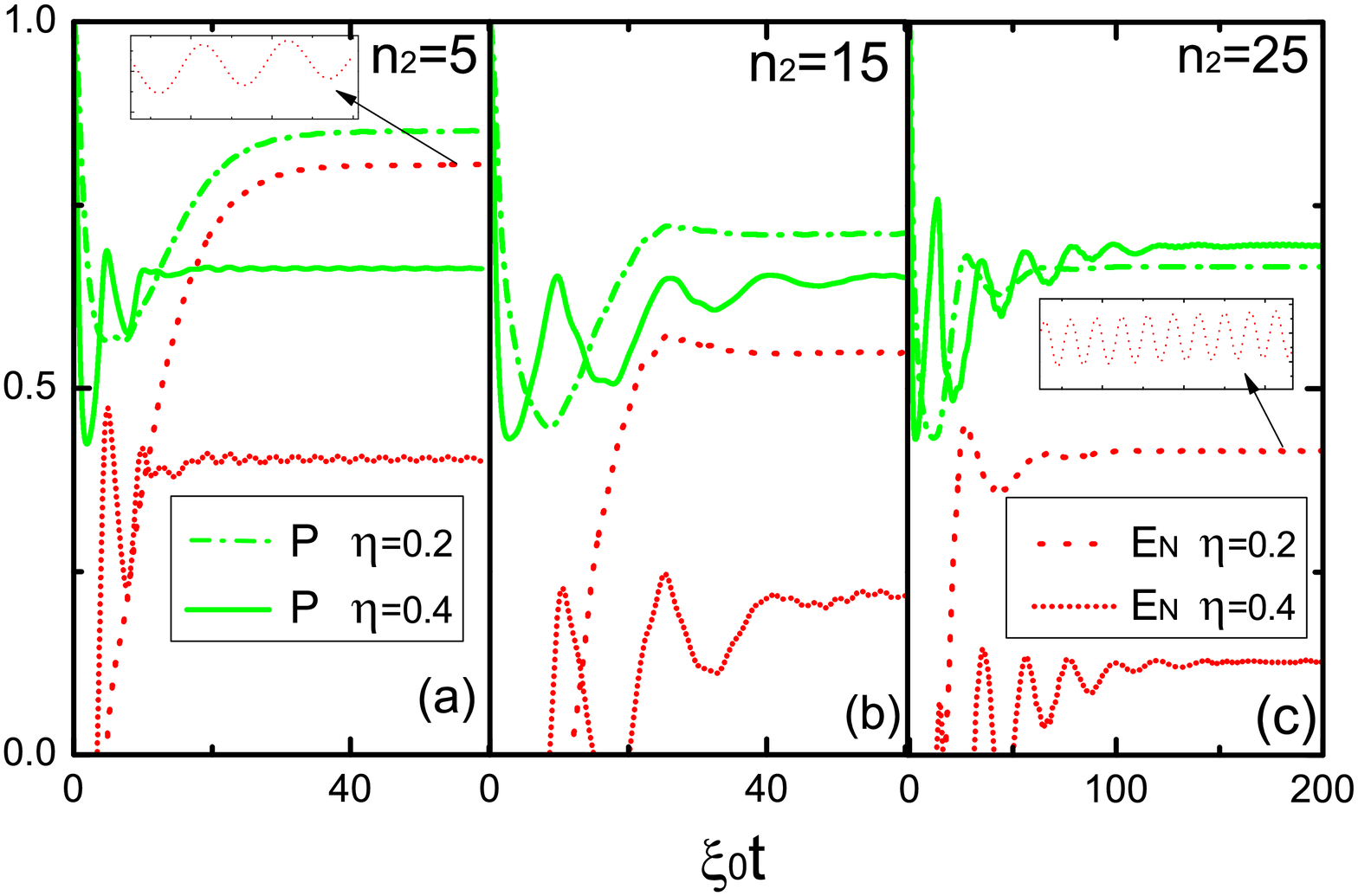}}} \caption{The
temporal behavior of the logarithmic negativity $E_N$ (red lines)
and the entanglement state purity $P$ (blue lines) for different
values of the coupling $\eta$ and the site $n_2$ the second cavity
located. The other parameters are the same as in Fig.~\ref{fig3}.}
\label{fig4}
\end{figure}

In Fig.~\ref{fig3}, the exact temporal evolution of the logarithmic
negativity $E_N$ and the purity $P$ of the two-mode cavity field are
plotted. From it, we see that the stationary logarithmic negativity
$E_N^s=r \simeq 1.0$ and the purity $P_s \simeq 1$ for all the cases
in which the first cavity siting at $n_1=1$ and the second cavity
siting at $n_2=5, 9, 15$ with the coupling $\eta=\xi_c/\xi_0=0.08,
0.05, 0.01$, respectively. It also shows that with the increase of
the distance between the two cavities, properly decreasing the
cavity-waveguide coupling $\eta$ leads to the same steady and pure
entangled state. This is because the oscillating profile of the
spectral density $J_{ii}(\omega)$, as shown in Fig.~\ref{fig2},
becomes narrower around the band center $\omega_0$ as the site $n_2$
increases.  This in turn requires longer time for achieving the
steady states as the coupling decreases. Thus, a proper choose of
the cavity-waveguide coupling strength can yield the same
entanglement degree and purity in the resonant case for the
different distances between the two cavities, as shown in
Fig.~\ref{fig3}.
\begin{figure}
\centerline{\scalebox{0.3}{\includegraphics{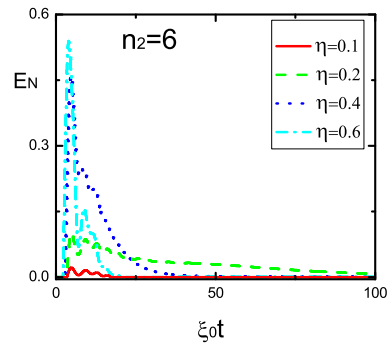}}} \caption{The
temporal behavior of the logarithmic negativity $E_N$. The other
parameters are the same as in Fig.~\ref{fig3}.} \label{fig5}
\end{figure}

In Fig.~\ref{fig4}, the entanglement and purity are plotted for the
enhanced cavity-waveguide couplings $\eta = 0.2$ and $0.4$, with the
different locations of the second cavity $n_2=5, 15$ and $25$. It
shows that as the cavity-waveguide coupling increases, the
entanglement decreases by accompanying with some oscillations in the
time evolution. The oscillation comes from the non-Markovian effect
due to the back-action between the cavity and the waveguide when the
cavity-waveguide coupling increases. Besides, one can also see that
the entanglement sudden death and sudden birth occurs in the
short-time regime with a relatively large cavity-waveguide coupling
($\eta=0.4$) and also a relatively long distance between the two
cavities, see Fig.~\ref{fig4}(b) and (c). In addition, as we see,
the resulting entangled states become usually a mixed state as the
cavity-waveguide coupling increases. By comparing
Fig.~\ref{fig4}~(a)-(c), one can also find that the entanglement
decreases as the increase of the distance between the cavities as
well as the cavity-waveguide coupling. Therefore, to maintain the
high entanglement for the two distant cavities, a small
cavity-waveguide coupling is more favorable in the resonant case.

Furthermore, as shown in Fig.~\ref{fig5} the entanglement appears in
the case of the first cavity siting at $n_1=1$ and the second cavity
siting at a even number of $n_2$. The exact numerical result shows
that the entanglement between the two cavities can be generated
under the relatively large coupling in a short time scale, but as
the time goes the entanglement soon decays to zero. This is
consistent with the analytical solution given by Eq.~(\ref{ssq}),
where it is pointed out that the steady-state entanglement in this
case can not exist if any one (or both) of the sites $n_1$ and $n_2$
of the two cavities is even.

\subsection{$|\omega_c-\omega_0|\ge2\xi_0$}
\begin{figure}
\centerline{\scalebox{0.4}{\includegraphics{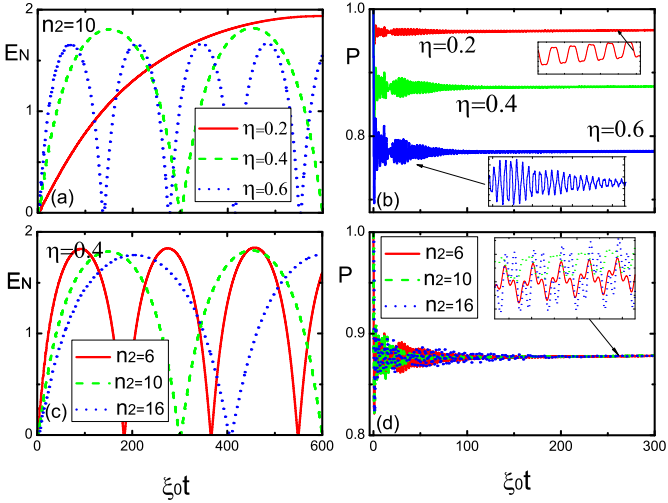}}} \caption{The
temporal behavior of the logarithmic negativity $E_N$ and the purity
$P$ for the different values of the coupling $\eta$ and the site
$n_2$, while the cavity frequency $\omega_c=1.2\omega_0$ which stays
the outside of the waveguide band. The other parameters are the same
as in Fig.~\ref{fig3}.} \label{fig6}
\end{figure}

To further investigate the controllability of the entanglement
generation of the two spatially-separated nanocavities, we consider
next the entanglement behavior for the cavity frequency $\omega_c$
outside the waveguide band, i.e., $|\omega_c-\omega_0|\ge2\xi_0$.
Let the first cavity site at $n_1=1$ and the second cavity locates
at different sites. The temporal behaviors of the entanglement and
the purity with different values of the cavity-waveguide coupling
are plotted in Fig.~\ref{fig6}. It shows that the entanglement
exhibits a very regular oscillation with a even stronger
entanglement degree $E_N^{\textrm{max}}>1$ for the input squeezing
$r=1$. For example, the maximal entanglement
$E_N^{\textrm{max}}\approx 1.94$ for the coupling $\eta=0.2$ in the
present case, see Fig.~\ref{fig6}(a). When the coupling is
increased, the entanglement degree oscillates faster but the maximal
entanglement is degraded a little bit only.

The above phenomenon can be understood as follows: in the weak
coupling region, the damping rates $\gamma_{ij}=0$ since the
spectral densities $J_{ij}(\omega_c)=0$ when the frequency
$\omega_c$ of the cavities lies outside the band of the waveguide.
However, the cavity frequency shift, $\delta \bm \omega= {\cal
P}\int d\omega \frac{\bm J(\omega)}{\omega-\omega_c}$ does not
vanish in this case. As a result, the reduced density matrix is
purely determined by the effective Hamiltonian
$H_{\textrm{eff}}=\sum_{i}(\omega_c+\delta\omega_{ii})a_i^\dag
a_{i}+\delta\omega_{12}(a_1^\dag a_2+a_2^\dag a_1)$. In other words,
an effective beam-splitter-type coupling (determined by the cross
frequency shift $\delta\omega_{12}$) between the two cavities is
induced by the waveguide, which results in the entanglement for the
separated
squeezing inputs. 
With the weak coupling solution of the propagating function
$\bm\mu_{bm}(t)$ given in the Appendix, we have explicitly
$\mu_{\textrm{bm},ii}=\cos(\delta\omega_{12}t)\textrm
e^{-i\tilde{\omega}_ct}$ and
$\mu_{\textrm{bm},12}=-i\sin{\delta\omega_{12}t}\textrm
e^{-i\tilde{\omega}_ct}$, where
$\tilde{\omega}_c=\omega_c+\delta\omega_{11}$ and
$\delta\omega_{22}\approx \delta\omega_{11} $. The average values in
Eq.~(\ref{sec}) reduce to $\langle a_i^\dag a_i\rangle=\sinh^2r$,
$\langle a_i^2\rangle=\sinh r\cosh r
\cos(2\delta\omega_{12}t)\textrm e^{-2i\tilde{\omega}_ct}$, $\langle
a_1a_2\rangle=-i\sinh r\cosh r \sin(2\delta\omega_{12}t)\textrm
e^{-2i\tilde{\omega}_ct}$, and $\langle a_1^\dag a_2\rangle=0$.
Thus, unlike the resonant case, the entanglement in this case is
purely determined by the cross frequency shift $\delta\omega_{12}$.
At the times when $\cos(2\delta\omega_{12}t)=0$, we have the nonzero
average values $\langle a_i^\dag a_i\rangle=\sinh^2r$ and $|\langle
a_1a_2\rangle|=\sinh r\cosh r$, which corresponds to a pure two-mode
squeezed vacuum state with the squeezing parameter $r$ \cite{gx}.
Accordingly, the entanglement degree at these times becomes optimal
with the maximal logarithmic negativity
\begin{align}
E_N^{\textrm max}=2r.
\end{align}
Fig.~\ref{fig6}(a) shows that the maximal entanglement approaches
the limit $2r$ when the coupling becomes sufficiently weak and the
weak coupling solution given in Appendix is almost exact. With the
cavity-waveguide coupling increase, the maximal entanglement is
decreased a little bit, as shown in Fig.~\ref{fig6}(a).

Fig.~\ref{fig6}(b) depicts the corresponding purity of the entangled
state. Since the damping rate vanishes in the weak coupling limit
when the frequency of the cavities lies outside the waveguide band,
the entangled state should be a pure state. Fig.~\ref{fig6}(b) shows
that for $\eta=0.2$, the exact numerical solution gives the purity
$P \approx 0.97$ which is consistent with the weak coupling
solution. When the cavity-waveguide coupling is increased, the
purity of the waveguide-generated entanglement is decreased.
Fig.~\ref{fig6}(c) and (d) show further the dependence of the
entanglement degree and the purity on the distance between the two
nanocavities. Similar behavior of the oscillating entanglement is
obtained as the cavity distance changes. This is due to the cross
frequency shift $\delta\omega_{12}$ again, which is determined by
the cross spectral density $J_{12}(\omega)$ that decreases slightly
as $n_2$ increases for the fixed $n_1=1$. Besides, it is also found
that the maximal entanglement and purity in the long-time region do
not change obviously with the changes of the distance between the
two nanocavities.
\begin{figure}
\centerline{\scalebox{0.32}{\includegraphics{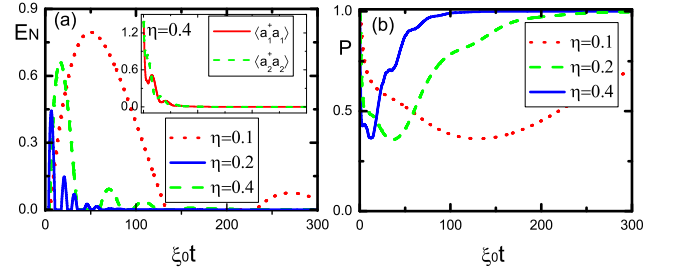}}} \caption{The
temporal behavior of the logarithmic negativity $E_N$ and the purity
$P$ for the cavity frequency within the waveguide band. Here we take
$\omega_c=1.03\omega_0$ and the site $n_2=5$. The insert in the
figure depicts the average photon numbers $\langle a_i^\dag
a_i\rangle$ of the two cavities. The other parameters are the same
as in Fig.~\ref{fig3}.} \label{fig7}
\end{figure}

\subsection{$|\omega_c-\omega_0|<2\xi_0$}
Finally, we shall consider the case that the cavity frequency
$\omega_c$ is not resonant with the band center but it still lies
inside the waveguide band. In Fig.7~(a), the entanglement and the
purity are plotted for the cavity sites $n_1=1$ and $n_2=5$ with the
cavity frequency $\omega_c=1.03\omega_0$ at which the spectral
density $J_{22}(\omega)=0$ for $\omega=\omega_c$ (see the red-dashed
line in Fig.~\ref{fig2}). In this case, the phenomenon of
entanglement sudden death and sudden birth occurs in the weak
coupling region. The existence of the entanglement sudden death and
sudden birth originates from the fact that the damping
$\gamma_{22}\approx 0$ and also $\gamma_{12}\approx 0$ for
$\omega_c=1.03\omega_0$ so that the entanglement is purely governed
by the cross frequency shift $\delta\omega_{12}$, which leads to the
entanglement oscillation. In the long time limit, the entanglement
is destroyed through the damping channel $\gamma_{11} \neq 0$. When
the cavity-waveguide coupling increases, the entanglement only
exists in a very short time and then quickly decay to zero
significantly, which is quite different from the resonant case shown
in Fig.~\ref{fig4}.  The insert in Fig.~\ref{fig7}(a) is the
intracavity average photon numbers $\langle a_j^\dag a_j\rangle$
which approach to zero in the long-time limit. It tells that the
cavity evolves asymptotically into a vacuum state. This is why the
entanglement in the steady-state vanishes. The purity shown in
Fig.~\ref{fig7}(b) approaches to one in the steady-state limit
because the steady state is just a trivial vacuum state.

\begin{figure}
\centerline{\scalebox{0.32}{\includegraphics{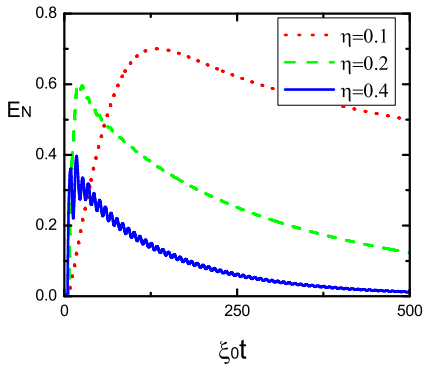}}} \caption{The
temporal behavior of the logarithmic negativity $E_N$ for the cavity
frequency $\omega_c=1.06\omega_0$ and the second cavity locating at
the site $n_2=5$. The other parameters are the same as in
Fig.~\ref{fig3}.} \label{fig8}
\end{figure}

In Fig.~\ref{fig8}, we plot the entanglement for the cavity
frequency $\omega_c=1.06\omega_0$, while the spectral density
$J_{ii}(\omega)\neq0~(i=1,2)$ at $\omega=\omega_c$. Then,  both the
decay channels $\gamma_{11}$ and $\gamma_{22}$ become active, and
the entanglement is a combination effect of the the non-zero cross
damping $\gamma_{12}$ and the non-zero cross frequency shift $\delta
\omega_{12}$. Compared to that in Fig.~\ref{fig7}(a), the
entanglement here is enhanced significantly in the long-time regime.
The phenomenon of the entanglement sudden death and sudden birth
disappears in this case. As a result, we see that the entanglement
generation can be easily controlled by changing the cavity frequency
within the band of the waveguide.

\section{Conclusions}
In conclusions, the generation of continuous variable entanglement
between two spatially-separated nanocavities mediated by a CROW in
photonic crystals is investigated. By solving the exact master
equation for the cavity system coupled to the waveguide as a
structured reservoir, the entanglement and the purity of the
two-mode cavity field is analyzed in details for the
initially-separated squeezing inputs. We found that the steady and
pure entangled state of the two distant cavities can be generated
for a weak cavity-waveguide coupling when the cavities is resonant
with the band center of the waveguide. It also shows that the strong
coupling of the cavities to the waveguide can lead to the phenomenon
of entanglement sudden death and sudden birth. When the cavity
frequencies lie outside the waveguide band, the optimal entanglement
can be achieved by the cross frequency shift between the two
cavities, which is induced by the waveguide. When the cavity
frequencies are not resonant with the band center but still within
the waveguide band, the entanglement can also exhibit the phenomenon
of sudden death and sudden birth even for a weak cavity-waveguide
coupling. By changing the cavity frequencies within the band of the
waveguide, the entanglement and the occurrence of entanglement
sudden death and birth can be easily controlled. These interesting
results show that the waveguide, as a controllable structured
reservoir, can be used to entangle efficiently the two
spatially-separated nanocavities. More importantly, only a weak
cavity-waveguide coupling is required for achieving the optimal
entanglement. We expect that these interesting features can be
realized in experiments and find its application in continuous
variable quantum communication networks.

\section*{Acknowledgment}
This work is supported by the National Science Council (NSC) of ROC
under Contract No. NSC-99-2112-M-006-008-MY3, the National Center
for Theoretical Science of NSC, the National Natural Science
Foundation of China (Grant Nos. 10804035, 60878004, and 11074087),
SRFDP (Grant Nos. 200805111014 and 200805110002), SDRF of CCNU
(Grant No. CCNU 09A01023), and the Natural Science Foundation of
Hubei Province.

\appendix*

\section{A weak-coupling analytical solution}

It should be pointed out that the master equation of Eq.~(\ref{me})
is exact, far beyond the Born-Markovian approximation and valid
for arbitrary cavity-waveguide coupling. The back-action between the
cavities and the waveguide is embedded into the time-dependent
coefficients $\omega_{ij}(t)$ and $\bm\gamma(t)$ of
Eq.~(\ref{tdcoefs}), which are in turn determined completely by the
propagating function $\bm \mu(t)$ of Eq.~(\ref{ut}). Generally, it
is not easy to obtain the analytical propagating function
$\bm\mu(t)$. However, for a weak coupling, the
analytical solution of the photon propagating function can be found as
\begin{align}
\label{mut} \bm\mu_{bm}(t)&=\textrm e^{-(\bm \gamma+i\bm{
\bar{\omega}})t},
\end{align}
where the damping rate becomes time-independent: $\bm\gamma=\bm
J(\omega_c)/2$, and the waveguide-induced renormalized frequency is
given by $\bm{\bar{\omega}}=\omega_cI+\bm{\delta\omega}$ with
\begin{eqnarray}
\bm{\delta\omega}=\mathcal{P}\int\frac{d\omega}{2\pi}\frac{\bm
J(\omega)}{\omega-\omega_c}.
\end{eqnarray}
Here $\mathcal{P}$ denotes the principle value of the integral. When
the cavity frequency $\omega_c$ is resonant with the band center
$\omega_0$ of the waveguide ($\omega_c=\omega_0$), the frequency
shift $\bm{\delta\omega}=0$, and the explicit propagating function
$\bm\mu_{bm}(t)$ in the weak-coupling limit is then given by
\begin{subequations}
\label{bm}
\begin{align}
\mu_{\textrm{bm},11}&=\frac{1}{\gamma_{11}+\gamma_{22}}[\gamma_{22}+\gamma_{11}\textrm
e^{-(\gamma_{11}+\gamma_{22})t}]\textrm e^{-i\omega_ct},
\\
\mu_{\textrm{bm},22}&=\frac{1}{\gamma_{11}+\gamma_{22}}[\gamma_{11}+\gamma_{22}\textrm
e^{-(\gamma_{11}+\gamma_{22})t}]\textrm e^{-i\omega_ct},
\\
\mu_{\textrm{bm},12}&=\mu_{\textrm{
bm},21}=-\frac{\gamma_{12}}{\gamma_{11}+\gamma_{22}}[1-\textrm
e^{-(\gamma_{11}+\gamma_{22})t}]\textrm e^{-i\omega_ct},
\end{align}
\end{subequations}
where the damping rates
$\gamma_{ii}=(\frac{2\xi_i^2}{\xi_0})\sin^2(\frac{n_i\pi}{2})$ and
the collective damping $\gamma_{12}=\sqrt{\gamma_{11}\gamma_{22}}$.

\end{document}